\documentclass[aps,prb,floatfix,epsfig,twocolumn,showpacs,preprintnumbers]{revtex4}
\UseRawInputEncoding
\usepackage{amsfonts}
\usepackage{amssymb}
\usepackage{graphicx,axodraw2}
\usepackage{amsmath}
\usepackage{tablefootnote}
\usepackage{setspace}
\begin{document}

\title{Electronic structure of LaNiO$_{2}$ and CaCuO$_{2}$ from self consistent vertex corrected GW approach}

\author{Andrey L. Kutepov\footnote{e-mail: akutepov@bnl.gov}}

\affiliation{Condensed Matter Physics and Materials Science Department, Brookhaven National Laboratory, Upton, NY 11973}

\begin{abstract}
Electronic structure of one of the nickelates (LaNiO$_{2}$) and one of the cuprates (CaCuO$_{2}$) is studied with three self consistent GW-based methods: scGW, sc(GW+Vertex), and quasiparticle self-consistent GW. Low energy features obtained in our study are in many respects similar to the features reported in previous DFT+DMFT studies. Consistent with the DFT+DMFT conclusion, we find LaNiO$_{2}$ as more correlated than CaCuO$_{2}$. However, correlation effects included in our study change the DFT Fermi surface near the $\Gamma$ point differently than it was reported in DMFT studies. Features which are a few electron-volts away from the Fermi level are broader in our calculations than in the DFT+DMFT which reflects the differences between the DFT and the GW methods. Our results are in qualitative agreement with previous G0W0 results, but the self-consistency brings in the quantitative differences. Generally, correlation effects are found to be sufficiently weak in both materials which allows one to use totally ab-initio diagrammatic approaches like sc(GW+Vertex) and to avoid the methods with adjustable parameters (DFT+U or DFT+DMFT). However, the possibility of some strong correlations at low energy which cannot be captured by perturbative methods cannot be completely excluded. For instance, differences in the Fermi surface should be resolved: experimental studies are necessary.
\end{abstract}

\maketitle


\section*{Introduction}
\label{intr}

Recent experimental discovery of the superconductivity in hole-doped NdNiO$_{2}$ generated renewed interest in nickelates. Particularly, the similarities and differences between nickelates and cuprates were studied intensely during the last couple of years. The study, obviously, is important as it can potentially reveal the features in the electronic structure which are responsible for the differences in superconducting properties. Very often, the study was concerned with LaNiO$_{2}$ and CaCuO$_{2}$ as simple representatives of both families of materials (nickelates and cuprates respectively). On the theoretical (calculational) side, the majority of the work was based on the Density Functional Theory\cite{prb_59_7901,prb_70_165109,prb_102_205130,prx_10_021061,prb_102_161118,prx_10_011024,prb_100_201106} (DFT),  or on the DFT plus Dynamical Field Theory (DFT+DMFT)\cite{prx_10_021061,prb_102_161118,prb_101_064513,commphys_3_84} calculations. Only one calculation based on the GW terminology applying its non-self-consistent version (G0W0) was published recently.\cite{prb_101_161102} From the methodological point of view it is important to mention also the application of the GW+DMFT approach to the related compound NdNiO$_{2}$.\cite{prx_10_041047} Before proceeding with the present work, let us capitalize briefly the most important for the present study results from other works.

At the DFT level, the principal difference between the two materials consists of the increased energy separation\cite{prb_70_165109} of the Ni 3d$_{x^{2}-y^{2}}$ orbitals from the O 2p orbitals in LaNiO$_{2}$ as compared to the corresponding separation of the Cu 3d$_{x^{2}-y^{2}}$ and the O 2p orbitals in CaCuO$_{2}$. Also, the 5d orbitals of La cross the Fermi level in LaNiO$_{2}$ and, therefore, are coupled with the Ni 3d$_{x^{2}-y^{2}}$ orbitals. At the DFT+DMFT level, Wang et al.\cite{prb_102_161118} studied two nickelates, SrNiO$_{2}$ and LaNiO$_{2}$. In their study, all Ni 3d orbitals were considered as correlated with the Hubbard U parameter 5 eV. Visual comparison of the electronic structure of LaNiO$_{2}$ obtained in [\onlinecite{prb_102_161118}] at the DFT and the DFT+DMFT levels (Fig. 2 in their work) does not show any qualitative differences on the scale of a few electron-volts. One can see the renormalization of bands only in the immediate vicinity of the Fermi level. Karp et al.\cite{prx_10_021061} used the DFT+DMFT theory to compare NdNiO$_{2}$ and CaCuO$_{2}$. Instead of considering all Ni(Cu) 3d states as correlated, authors of Ref. [\onlinecite{prx_10_021061}] performed two types of the DFT+DMFT calculations: one with only Ni(Cu) 3d$_{x^{2}-y^{2}}$ as correlated orbital, and the second with Ni(Cu) 3d$_{x^{2}-y^{2}}$ and 3d$_{3z^{2}-r^{2}}$ as correlated. Hubbard parameter U was, correspondingly, increased from 3.1 eV in the first type of calculations to 7 eV in the second one. The important conclusion from this work is that nickelates are more correlated than cuprates (see Fig. 2 in Ref. [\onlinecite{prx_10_021061}]). Also, authors place nickelates in the same charge-transfer category of materials as the cuprates despite the larger separation between the d$_{x^{2}-y^{2}}$ and the O 2p states in the nickelates. One more important difference is that the rare earth d states appear in both the addition and removal spectra in nickelates which is a sign of their hybridization with Ni 3d states.

Cited above DFT and DFT+DMFT works provide insightful information on the materials. One can point out, however, that there are certain issues, particularly with the DFT+DMFT, which can affect the robustness of the conclusions. Firstly, the DFT+DMFT results depend on the choice of the U parameter. In this respect, the choice of U in Refs.[\onlinecite{prx_10_021061}] and [\onlinecite{prb_102_161118}] seems to be inconsistent. In the first work the U parameter was 3.1 eV for 1 correlated orbital and 7 eV for two correlated orbitals. The more orbitals we consider as correlated, the larger U should be because the screening by the rest of the (uncorrelated) orbitals is reduced. However, in Ref. [\onlinecite{prb_102_161118}], where all five Ni 3d orbitals were correlated, U value was only 5 eV. Secondly, because of the apparent importance of the energy separation between the Ni(Cu) 3d and the O 2p levels and of the degree of the hybridization between the Ni 3d and the La 5d, neglect by the inter-site (non-local) components of self energy in both the DFT and the DFT+DMFT studies seems to be highly questionable when considering relative positioning of the Ni(Cu) 3d and O 2p states, and Ni 3d and La 5d. Thirdly, low energy physics (immediate vicinity of the Fermi level) which is the principal goal of the DFT+DMFT studies can most likely be also affected by the effects not included in the DFT+DMFT: electron-phonon interaction, the same non-local self energy effects, and by the frequency dependence of the effective interaction. Therefore, the conclusions might change when all important contributions are properly taken into account.

Thus, it seems to be important and interesting to also apply other methods which include correlation effects and which are free of at least some of the mentioned issues of the DFT+DMFT. In this respect, the work by Olevano et al., [\onlinecite{prb_101_161102}], represents an important step. In their work, the non self-consistent GW approximation (G0W0) was used to study the electronic structure of LaNiO$_{2}$. G0W0 represents only the first term in the expansion of self energy but includes all non-local physics on the same footing as the local one. Plus, it has no adjustable parameters and it considers full frequency dependent effective interaction. Authors of Ref. [\onlinecite{prb_101_161102}] have shown that the La 4f states undergo 2 eV upward shift with respect to their DFT position, whereas the O 2p states are pulled down by 1.5 eV. Thus, they stress the importance of the non-local physics in this compound. As a drawback of the G0W0 approximation one can consider its obvious dependence on the starting point (because of the lack of self consistency). G0W0 relies on the assumption that GW wave functions 
are similar to the DFT wave functions (if the DFT is used as a starting point). This assumption works well in simple semiconductors (like Si or LiF) but can be seriously questioned in more complicated materials. For instance, the G0W0 (with the DFT as a stating point) applied to the monoclinic M1 phase of VO$_{2}$ results in a metal (similar to the DFT) whereas it is an insulator in experiments.\cite{prl_99_266402} Only the self-consistent quasiparticle GW calculation provides correct insulating state.\cite{prr_2_023076} With this consideration, it is clear that the self consistent and based on GW calculations can provide essential new information on the differences in the electronic structure of LaNiO$_{2}$ and CaCuO$_{2}$.

The principal goal of this work is, therefore, to apply self-consistent GW method to the representatives of nickelates and cuprates. We also apply self-consistent GW+Vertex approach to determine the strength of the correlation effects beyond GW approximation. Plus, we apply self-consistent quasiparticle GW approximation (QSGW) as it stresses the importance of the Ward Identity (WI) in the limit of low frequency and low momenta but neglects the dynamical effects (frequency dependence) in self energy. The scGW, on the other hand, treats the high frequency part of self-energy on the same footing as the low frequency part but neglects the WI altogether.

The paper begins with a brief discussion of the distinctive features of the methods used in this work and setup parameters for the calculations (the first section). The second section provides the
results obtained and a discussion. The conclusions are given afterwords. Finally, three sections in the Appendix provide supporting for the main text information.

\section*{Methods and calculation setups}\label{meth}

All calculations in this work were performed using the code FlapwMBPT.\cite{flapwmbpt} For the DFT calculations, we used the local density approximation (LDA) as parametrized by Perdew and Wang.\cite{prb_45_13244} Recently, a number of improvements in the quality of the basis set in the FlapwMBPT code have been implemented\cite{prb_103_165101} which allowed, for instance, more accurate evaluation of the atomic forces.\cite{jcm_press} Our scGW and sc(GW+Vertex) calculations are based on the L. Hedin's theory.\cite{pr_139_A796} They can also be defined using $\Psi$-functional formalism of Almbladh et al.\cite{ijmpb_13_535} As it is shown in Ref. [\onlinecite{ijmpb_13_535}], $\Psi$-functional can be constructed starting from Luttinger-Ward $\Phi$-functional\cite{pr_118_1417} and using screened Coulomb interaction W instead of bare Coulomb interaction V as an independent variable (besides Green's function G). It is defined by the following expression:

\begin{equation}\label{psi}
\Psi[G,W]=\Phi[G,V]-\frac{1}{2}Tr [PW-ln(1+PW)],
\end{equation}
where $P$ is irreducible polarizability. In materials science, $\Psi$-functional is more convenient than the $\Phi$-functional of Luttinger and Ward. The first reason is connected to the infinite range of the bare Coulomb interaction which makes the screened interaction W a much more suitable quantity than the bare interaction
V. The second reason is the simplicity of $\Psi$-functional. For instance, at the level of GW approximation, $\Phi$-functional is represented by an infinite sequence of ring diagrams whereas $\Psi$-functional is represented by just one diagram (first diagram in Fig. \ref{diag_Psi}). In this work, the simplest approximation for $\Psi$-functional which includes vertex corrections has been adapted (Fig. \ref{diag_Psi}). As it was already mentioned, the first diagram in Fig. \ref{diag_Psi} corresponds to GW approximation whereas the second one represents the first order vertex correction.

\begin{figure}[t]
\begin{center}\begin{axopicture}(200,56)(0,0)
\SetPFont{Arial-bold}{28}
\SetWidth{0.8}
\Text(10,10)[l]{$\Psi$ =}
\Text(35,10)[l]{$-\frac{1}{2}$}
\GCirc(75,10){20}{1}
\Photon(55,10)(95,10){2}{5.5}
\Text(110,10)[l]{+}
\Text(128,10)[l]{$\frac{1}{4}$}
\GCirc(160,10){20}{1}
\Photon(160,-10)(160,30){2}{5.5}
\Photon(140,10)(180,10){2}{5.5}
\end{axopicture}
\end{center}
\caption{Diagrammatic representation of $\Psi$-functional which includes the simplest non-trivial vertex.}
\label{diag_Psi}
\end{figure}

\begin{figure}[t]
\begin{center}\begin{axopicture}(200,56)(0,0)
\SetPFont{Arial-bold}{28}
\SetWidth{0.8}
\Text(10,10)[l]{$P$  =}
\Photon(48,10)(55,10){2}{2.5}
\GCirc(75,10){20}{1}
\Photon(95,10)(102,10){2}{2.5}
\Text(120,10)[l]{$-$}
\Photon(143,10)(150,10){2}{2.5}
\GCirc(170,10){20}{1}
\Photon(190,10)(197,10){2}{2.5}
\Photon(170,-10)(170,30){2}{5.5}
\end{axopicture}
\end{center}
\caption{Diagrammatic representation of irreducible polarizability in the simplest vertex corrected scheme.}
\label{diag_P}
\end{figure}

\begin{figure}[t]
\begin{center}\begin{axopicture}(200,56)(0,0)
\SetPFont{Arial-bold}{28}
\SetWidth{0.8}
\Text(10,10)[l]{$\Sigma$  = -}
\Line(38,10)(112,10)
\PhotonArc(75,10)(30,0,180){2}{8.5}
\Text(120,10)[l]{$+$}
\Line(133,10)(207,10)
\PhotonArc(160,10)(20,0,180){2}{6.5}
\PhotonArc(180,10)(20,180,360){2}{6.5}
\end{axopicture}
\end{center}
\caption{Diagrammatic representation of self energy in the simplest vertex corrected scheme.}
\label{diag_S}
\end{figure}

Diagrammatic representations for irreducible polarizability (Fig. \ref{diag_P}) and for self energy (Fig. \ref{diag_S}) follow from the chosen approximation for $\Psi$-functional. The set of diagrams for polarizability and self energy shown in Figs. \ref{diag_P} and \ref{diag_S} corresponds to the scheme B introduced earlier in Ref. [\onlinecite{prb_94_155101}]. In order to make notations more self-explaining, here we introduce another abbreviation. Following the convention for the GW approach which corresponds to the lines of the GW diagram (first diagram in Fig. \ref{diag_S}), we will use the term sc(GW+G3W2) (instead of sc(GW+Vertex) or "scheme B") which corresponds to all diagrams in Fig. \ref{diag_S}. Specific diagrammatic representation of polarizability defines the approximation for screening. Thus, from Fig. \ref{diag_P} we can state that in sc(GW+G3W2) the screening is defined by one-loop diagram (Random Phase Approximation, RPA) plus first order electron-hole interaction diagram. scGW includes only the RPA part. Technical details of the GW part were described in Refs. [\onlinecite{prb_85_155129,cpc_219_407}]. Numerical algorithm for the evaluation of first order polarizability was the same in this study as described in details in Ref. [\onlinecite{prb_94_155101}]. For the evaluation of second order self-energy, however, more efficient algorithm (as compared to the one described in [\onlinecite{prb_94_155101}]) is used. The brief account of the details of this new algorithm can be found in Appendix. The diagrammatic (GW and G3W2) parts of the FlapwMBPT code take full advantage of the fact that certain diagrams can more efficiently be evaluated in reciprocal (and frequency) space whereas other diagrams are easier to evaluate in real (and time) space. As a result, the GW part of the code scales as $N_{k}N_{\omega}N^{3}_{b}$ where $N_{k}$ is the number of \textbf{k}-points in the Brillouin zone, $N_{\omega}$ is the number of Matsubara frequencies, and $N_{b}$ stands for the size of the basis set. The vertex part of the code scales as $N^{2}_{k}N^{2}_{\omega}N^{4}_{b}$. For comparison, if one uses naive (all in reciprocal space and frequency) implementation then the GW part scales as $N^{2}_{k}N^{2}_{\omega}N^{4}_{b}$ (i.e. exactly as the vertex part when the implementation is efficient), and the vertex part scales as $N^{3}_{k}N^{3}_{\omega}N^{5}_{b}$. Besides of the efficiency of implementation, we have to mention two more factors which make the use of the diagrams beyond GW feasible. First is the fact that the higher order diagrams converge much faster than the GW diagram with respect to the basis set size and to the number of \textbf{k}-points.\cite{prb_94_155101,prb_95_195120} Second is that the higher order diagrams are very well suited for massive parallelization.

scGW has a certain advantage as compared to non self consistent (one shot) G0W0 approach: there is no dependence on the starting point in scGW. Also, being based on the functional formalism, it allows (at least in principle) the direct way to evaluate total energies.\cite{prb_57_2108,prb_80_041103,jcm_29_465503} However, from the purely theoretical point of view, scGW has certain issues which one can relate to rather "nonsymmetric" dressing of the Green function during the self-consistency course: adding more and more self-consistency diagrams while retaining at each iteration only the lowest order skeleton diagram for polarizability and for self energy. This "nonsymmetric" dressing results in, for instance, incorrect long wave limit of polarizability. There is quite a number of documented limitations of the approach: too big (as compared to the correct result) bandwidth in electron gas\cite{prb_57_2108} and in alkali metals\cite{prl_81_1662}, absence of satellites in electron gas\cite{prb_57_2108}, overestimation of the band gap in simple semiconductors.\cite{prl_81_1662,prb_98_155143} In order to "defend" scGW a bit, one can observe that the above listed limitations pertain mostly to the materials where non local physics is prevalent (electron gas, alkali metals, sp semiconductors). There is only limited number of scGW applications to the realistic materials where local effects are the most important or, at least, contribute considerably to observable properties. Existing applications, however, are not as conclusive as in the case of simple materials. Just to name a few, one can point out that scGW overestimates magnetic moment in iron\cite{jcm_29_465503} and band gap in NiO\cite{arx_2106_03800}. Also, in SrVO$_{3}$, there is an indication of worsening of the calculated spectra when go from G0W0 to scGW.\cite{prb_94_201106} However, scGW in Ref. [\onlinecite{prb_94_201106}] was implemented for the basis set of rather small size (only $t2g$ orbitals), which makes the conclusion though plausible but not very convincing. On the other hand, scGW describes perfectly well the experimental photoemission spectrum of metal americium\cite{prb_85_155129} whereas G0W0 fails completely. Also, applications of scGW are rather popular in atomic and molecular physics\cite{epl_76_298,jcp_130_114105,prb_86_081102}  which supports an idea that in the "finite systems" world, scGW has certain merits.

sc(GW+G3W2) adds skeleton diagrams of the next order (as compared to scGW) to both polarizability and self energy. Therefore, the problems occuring because of the above described "nonsymmetric" dressing of Green's function should be less dramatic. From this point of view, one can expect sc(GW+G3W2) to be more accurate than scGW. Indeed, there is noticeable improvement in the calculated bandwidth of the electron gas\cite{prb_96_035108} and alkali metals\cite{prb_94_155101}. Improvements in the calculated band gap of sp semiconductors are especially remarkable\cite{prb_95_195120}. In the case of simple semiconductors, sc(GW+G3W2) not only considerably outperforms scGW and QSGW (see introduction below), but also is better than G0W0 in most of the cases. For more complicated materials, one can point out recent calculation of the band gap in NiO\cite{arx_2106_03800} where sc(GW+G3W2) resulted in almost perfect reproduction of the experimental gap whereas scGW overestimated it by about 25\%. Also, the improvement in the calculated band gap of the van der Waals ferromagnet CrI$_{3}$ is considerable.\cite{arx_2105_07798} Of course, one cannot expect that sc(GW+G3W2) will be considerably better than scGW in the case of really strongly correlated materials, i.e. where non-perturbative treatment is necessary.

We also use the quasiparticle self consistent GW (QSGW) approach. Similar to the scGW and the sc(GW+G3W2) approaches, it is based on the finite temperature (Matsubara) formalism and in this respect it is different from the well known QSGW implementation by Kotani et al.\cite{prb_76_165106} Quasiparticle approximation includes linearization of self energy near zero frequency (see for details Refs. [\onlinecite{prb_85_155129,cpc_219_407}]) and, therefore, the method is reliable only not very far from the Fermi level - usually within a few electron-volts. The approach adopted by Kotani et al. in Ref. [\onlinecite{prb_76_165106}] uses specially designed procedure of averaging of non-diagonal elements of self energy for each quasiparticle state instead of the linearization near the chemical potential. This fact, presumably, should make the approach of Kotani et al. more accurate in broad energy range than QSGW used in this study. However, for the energies not far from the chemical potential (the range of interest in this work) two types of QSGW are quite similar. The differences, in fact, are mostly related to the differences in basis sets and in the degree of convergence.\cite{prb_95_195120} In both variants of QSGW, the effective self energy is static (frequency independent, see App. \ref{sig_qsgw}) and the method is not diagrammatic. Special (or rather manual) construction of the effective self energy breaks relation to the $\Psi$-functional. However, as it was explained by Kotani et al.\cite{prb_76_165106} QSGW satisfies the zero frequency and long wave limit of the Ward Identity because of the so called Z-factor cancellation. This fact makes it often quite accurate, especially in simple metals and semiconductors where the above mentioned limit is important. Band gaps, calculated with the QSGW, for instance, are usually more accurate than the ones calculated with the scGW.\cite{prl_99_246403, prb_95_195120,prb_98_155143} In more complicated solids (especially where d or f electrons play an important role) the QSGW approach is not necessary better than the scGW: the frequency dependence of self energy could be more important than the zero frequency+momentum limit of the WI. Good example is the metal americium, where both the DFT and the QSGW fail to describe the experimentally determined\cite{prl_52_1834} position of the occupied 5f$_{5/2}$ states whereas scGW describes them very well.\cite{prb_85_155129} For simple (sp) semiconductors with large band gap (C, MgO, LiF, NaCl) scGW outperforms QSGW\cite{prb_98_155143,prb_95_195120} (not considerably though). Also, as it seems\cite{arx_2105_07798}, scGW is slightly more accurate than QSGW in the case of CrI$_{3}$. Additional insight into the differences between the approximate methods of this work is provided in Appendix \ref{Im_W}. Considering their differences, the three approaches (scGW, sc(GW+G3W2), and QSGW) represent a good set of methods to study new materials.

Our algorithm for the analytical continuation of self energy which was needed, for instance, to plot Figs. \ref{pdos_ca} and \ref{pdos_la} is based on the Ref. [\onlinecite{jltp_29_179}] and it is described in the Appendix of Ref. [\onlinecite{cpc_257_107502}]. The band plotting associated with the scGW/sc(GW+G3W2) approach (see Fig.\ref{gzr_bnd}) needs some additional clarification. Strictly speaking, one-electron features (band dispersions) in these two approaches should be obtained as the peak positions of the $\mathbf{k}$-resolved spectral functions. Evaluation of spectral functions includes the analytical continuation of the correlation part of self energy from the imaginary to the real frequency axis. However, as it was demonstrated in Refs. [\onlinecite{prb_85_155129,cpc_257_107502}], the peak positions of the spectral function near the chemical potential can often be accurately reproduced by a simplified procedure. This procedure involves the linearization of the frequency dependence of self energy near the chemical potential and, consequently, results in the effective one-electron energies (see details in Appendix of the Ref. [\onlinecite{cpc_257_107502}]). The one-electron energies, such obtained, can obviously be used for the band plotting purposes.

Let us now specify the setup parameters used in the calculations. In order to make presentation more compact, principal structural parameters for the studied solids have been collected in Table \ref{list_s} and the most important set up parameters have been collected in Table \ref{setup_s}. All calculations have been performed for the electronic temperature $600K$. As the long range magnetic order has not yet been found in LaNiO$_{2}$, all calculations were non-magnetic for simplicity. The DFT, scGW, QSGW, and the GW part in the sc(GW+G3W2) calculations were performed with the $6\times 6\times 6$ mesh of \textbf{k}-points in the Brillouin zone. 300 band states were used to expand Green's function and self energy. The convergence which the above parameters provide was checked by doing calculations with $4\times 4\times 4$ mesh of \textbf{k}-points and with smaller number of bands for the GW part. From the analysis we conclude that further increase in the number of \textbf{k}-points and bands should not change the effective band energies near the Fermi level (Fig. \ref{gzr_bnd}) by more than 5\% which is sufficient for the comparison of methods. The diagrams beyond the GW approximation were evaluated using $3\times 3\times 3$ mesh of \textbf{k}-points in the Brillouin zone and with about 26 bands (closest to the Fermi level). With the above mentioned faster convergence of the higher order diagrams with respect to these parameters, this choice represented a reasonable compromise between the accuracy and the computational cost. Similar to GW part, the convergence was checked by doing calculations with smaller number of \textbf{k}-points ($2\times 2\times 2$) and of bands (10-22 instead of final 26). We estimate the error of the vertex part (i.e. the difference between sc(GW+G3W2) and scGW results) to be about or less than 10-15\%. Again, from Fig. \ref{gzr_bnd} one can see that the above difference is pretty small itself so that if it changes by 10-15\% the conclusions will be the same.

\section*{Results}
\label{res}

\begin{table}[t]
\caption{Structural parameters of the solids studied in this work. Lattice parameters are in Angstroms, MT radii are in atomic units (1 Bohr radius), and atomic positions are given relative to the three primitive translation vectors.} \label{list_s}
\small
\begin{center}
\begin{tabular}{@{}c c c c c c} &Space&&&Atomic&\\
Solid &group&a&c&positions&$R_{MT}$\\
\hline\hline
CaCuO$_{2}$&123 &3.86 &3.20&Ca: 0;0;0  &2.032\\
& & & &Cu: 1/2;1/2;1/2  &2.032\\
& & & &O: 1/2;0;1/2  &1.563\\
LaNiO$_{2}$&123 &3.966 &3.376&La: 0;0;0  &2.087\\
& & & &Ni: 1/2;1/2;1/2  &2.087\\
& & & &O: 1/2;0;1/2  &1.606\\
\end{tabular}
\end{center}
\end{table}

\begin{table}[b]
\caption{Principal setup parameters of the studied solids are given. The following abbreviations are introduced: $\Psi$ is for wave functions, $\rho$ is for the electronic density, $V$ is for Kohn-Sham potential, and PB is for the product basis.} \label{setup_s}
\small
\begin{center}
\begin{tabular}{@{}c c c c c c} &Core&&$L_{max}$&$L_{max}$&\\
Solid &states&Semicore&$\Psi/\rho,V$&PB & $RK_{max}$ \\
\hline\hline
CaCuO$_{2}$&Ca: [Ne]& 3s,3p&6/6&6&8.0  \\
& Cu: [Ne]& 3s,3p&6/6&6&  \\
& O: [He]& 2s&5/5&5&  \\
LaNiO$_{2}$&La: [Ar]3d& 4s,4p,4d,5s,5p&6/6&6&8.0  \\
& Ni: [Ne]& 3s,3p&6/6&6&  \\
& O: [He]& 2s&5/5&5&  \\
\end{tabular}
\end{center}
\end{table}

\begin{figure*}[t]       
    \fbox{\includegraphics[width=6.5 cm]{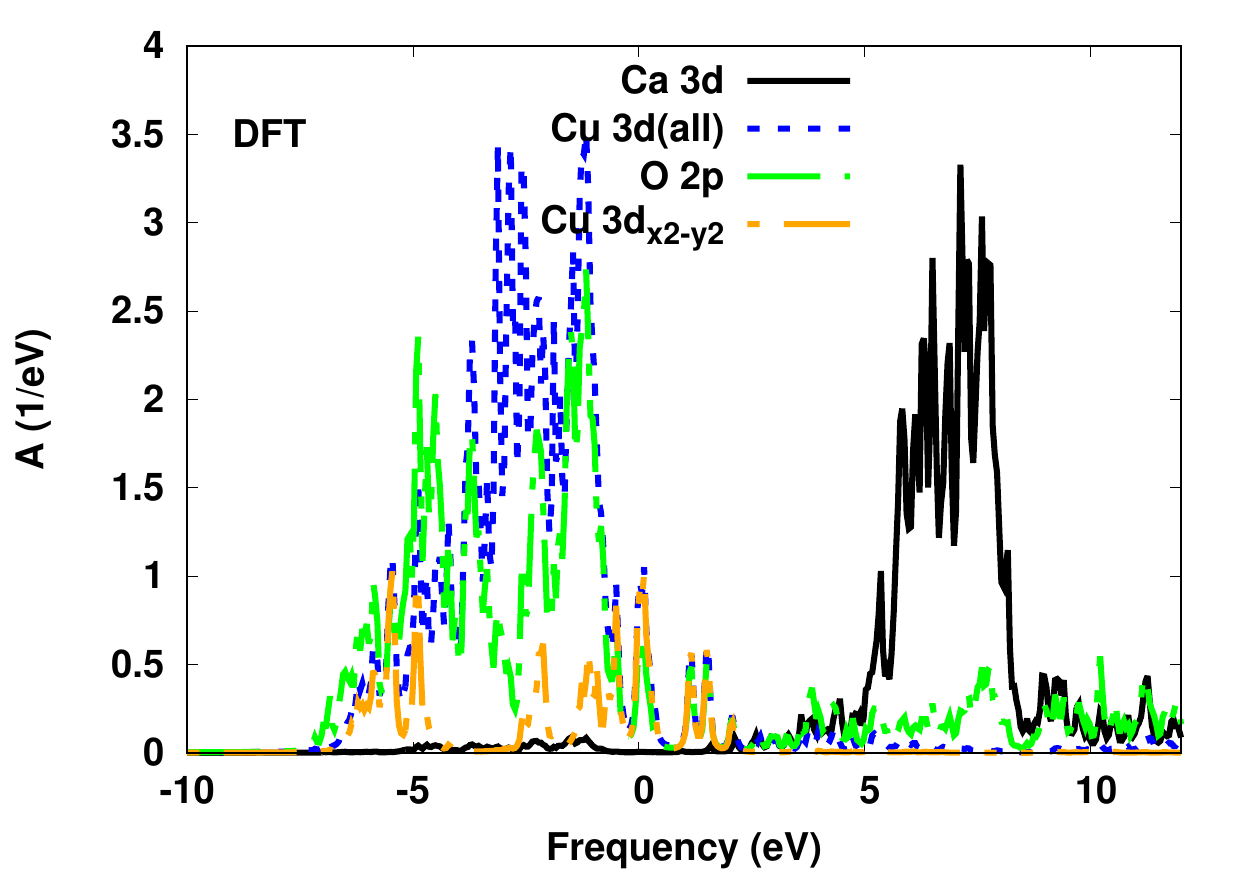}}   
    \hspace{0.02 cm}
    \fbox{\includegraphics[width=6.5 cm]{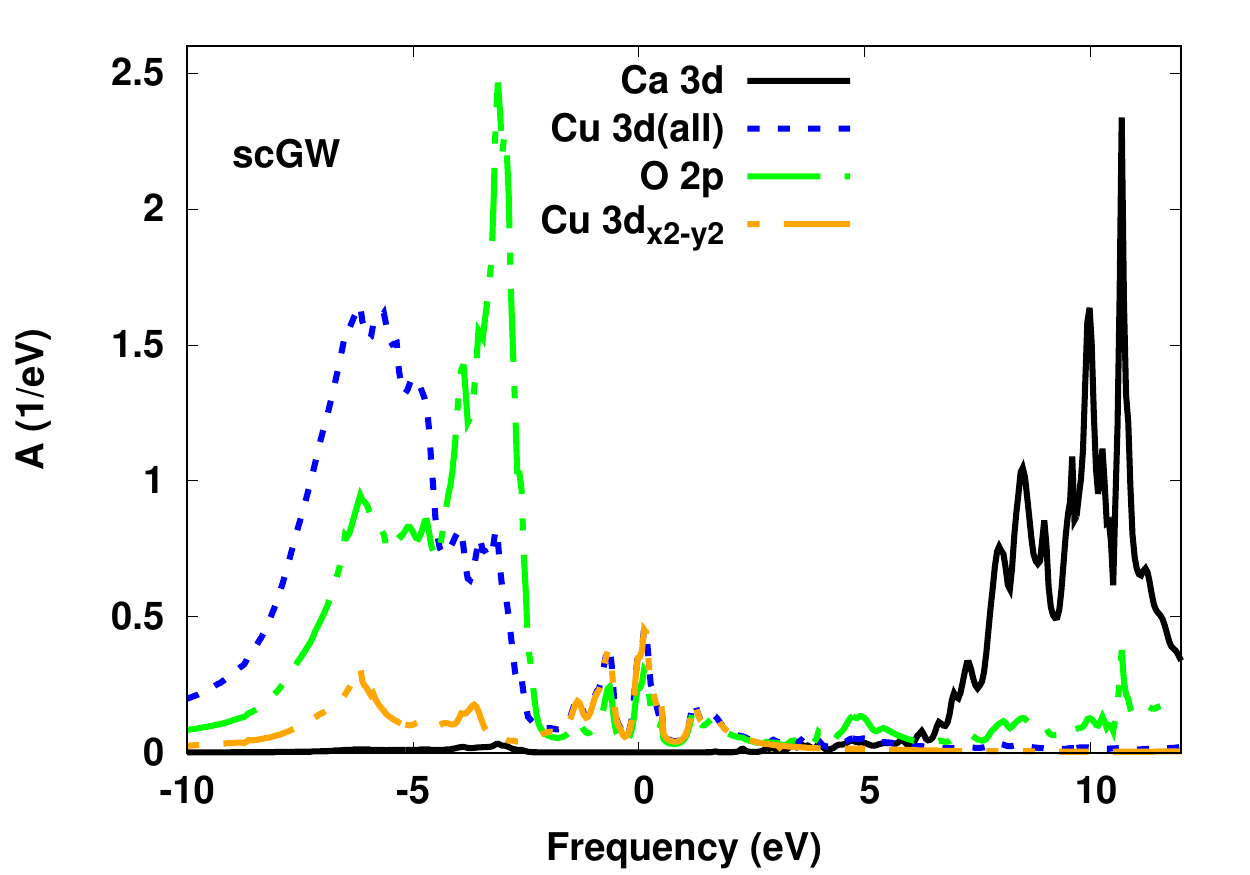}}  
    \hspace{0.02 cm}
    \fbox{\includegraphics[width=6.5 cm]{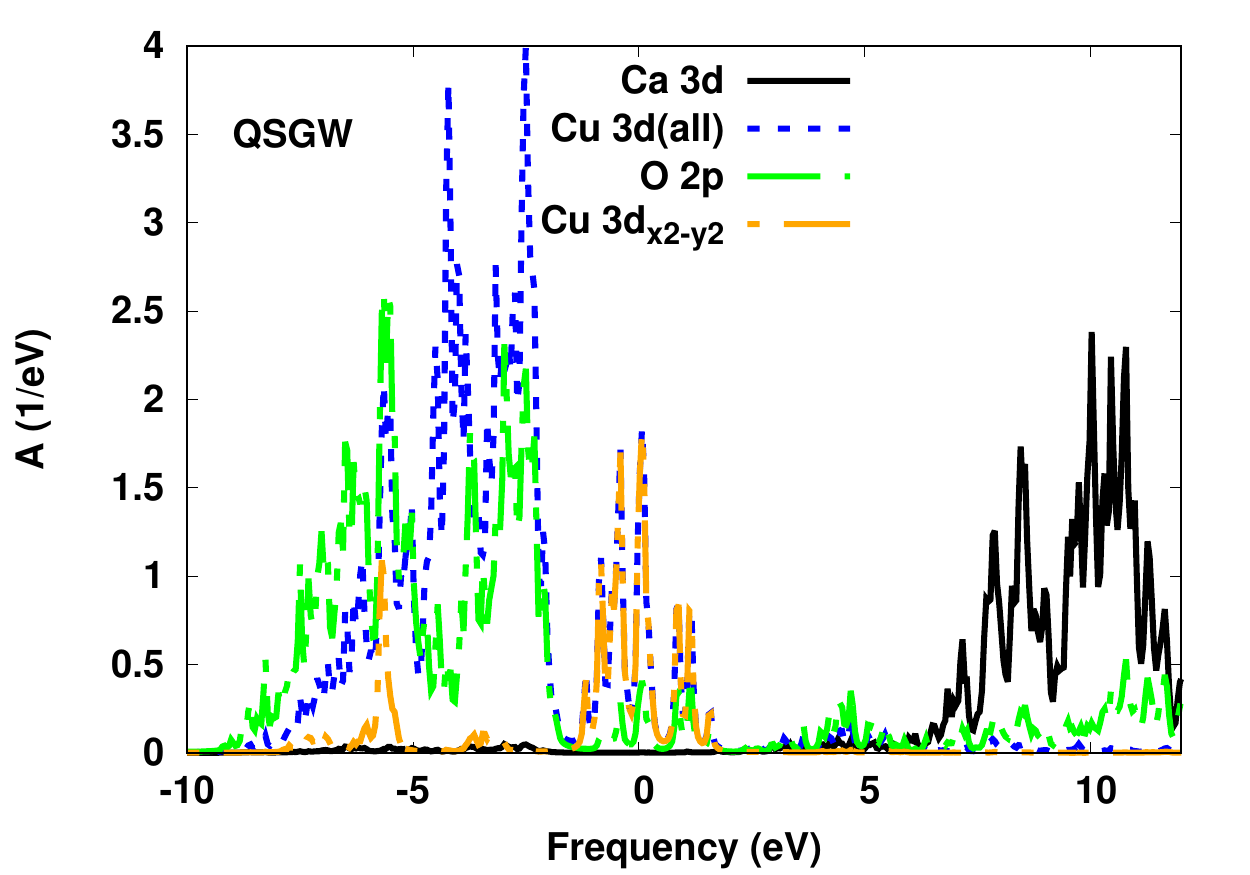}}  
    \hspace{0.02 cm}
    \fbox{\includegraphics[width=6.5 cm]{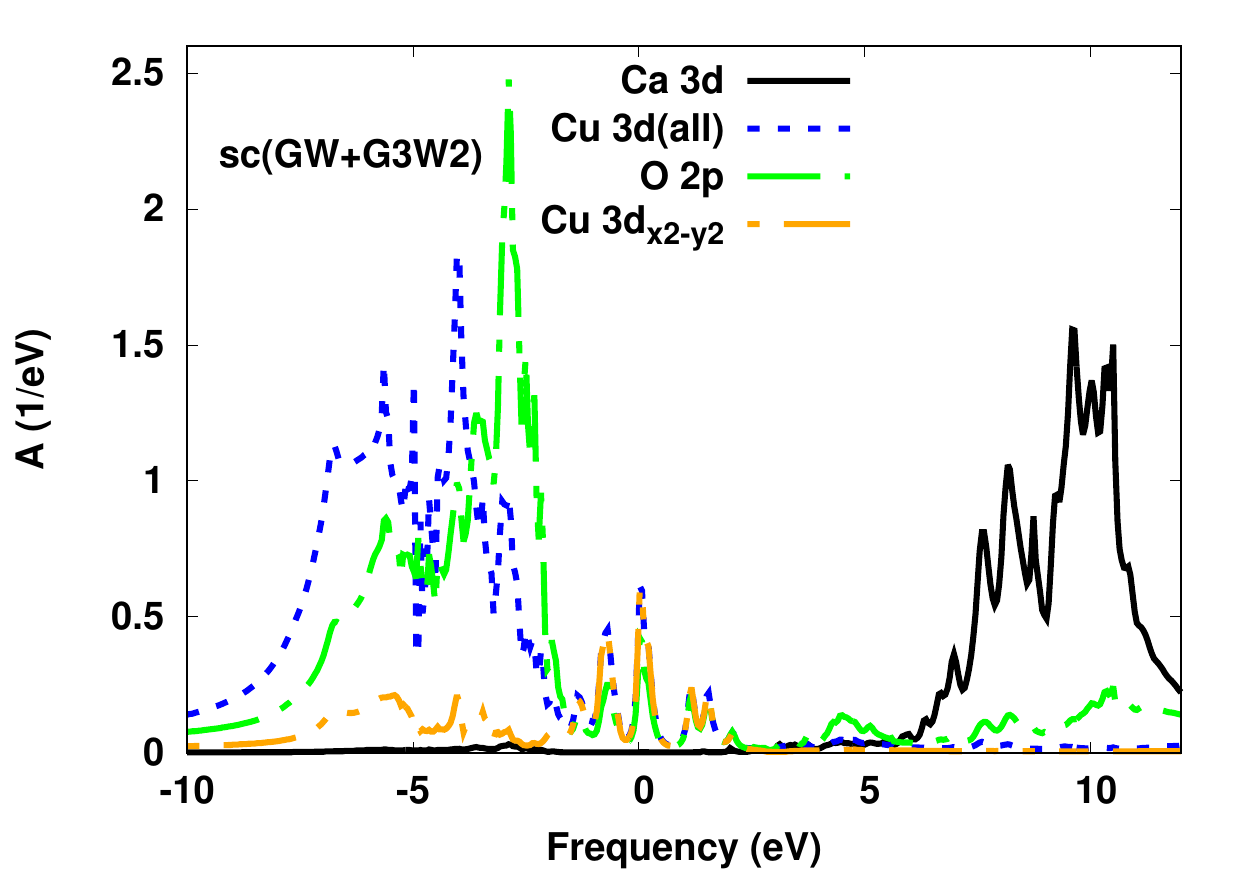}}
    \caption{Partial (atom and orbital resolved) spectral functions of CaCuO$_{2}$.}
    \label{pdos_ca}
\end{figure*}

\begin{figure*}[t]       
    \fbox{\includegraphics[width=6.5 cm]{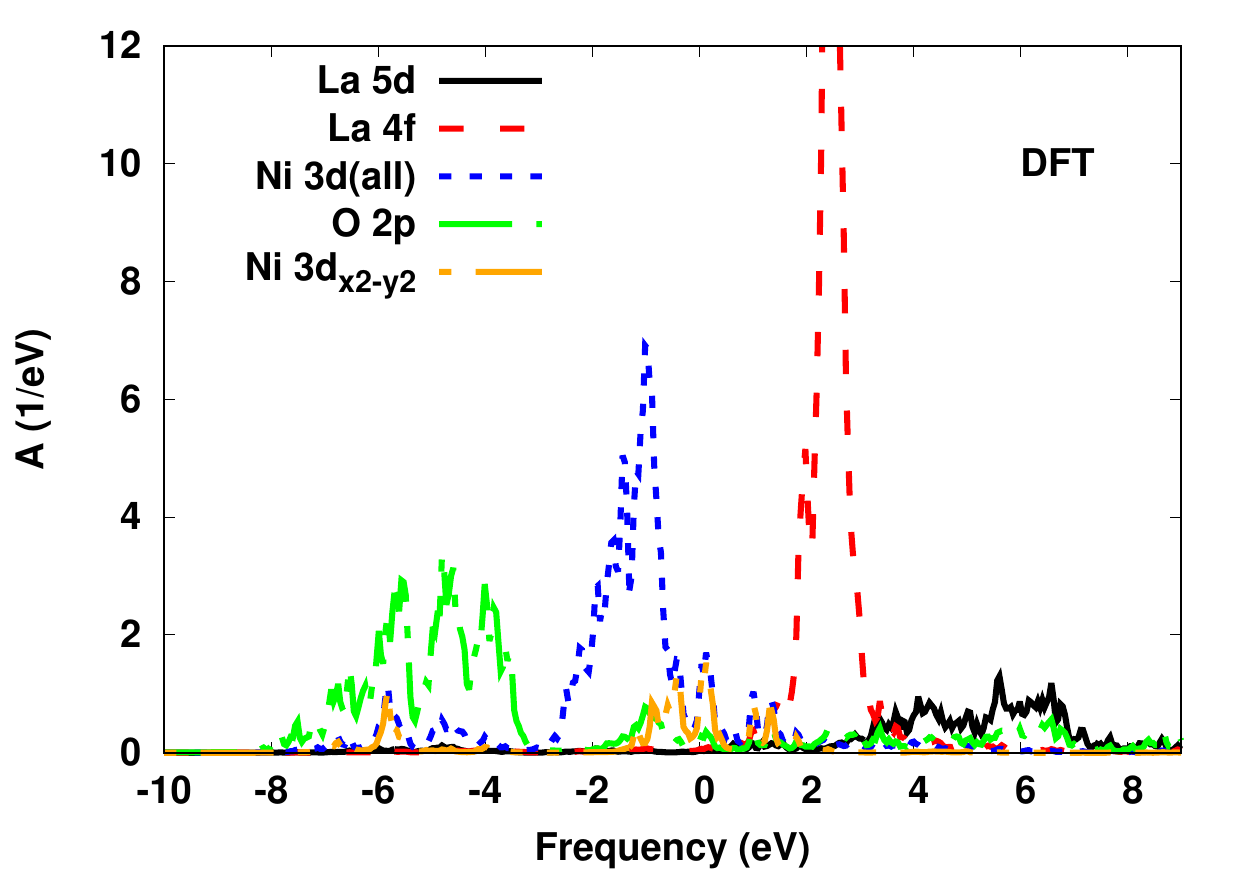}}   
    \hspace{0.02 cm}
    \fbox{\includegraphics[width=6.5 cm]{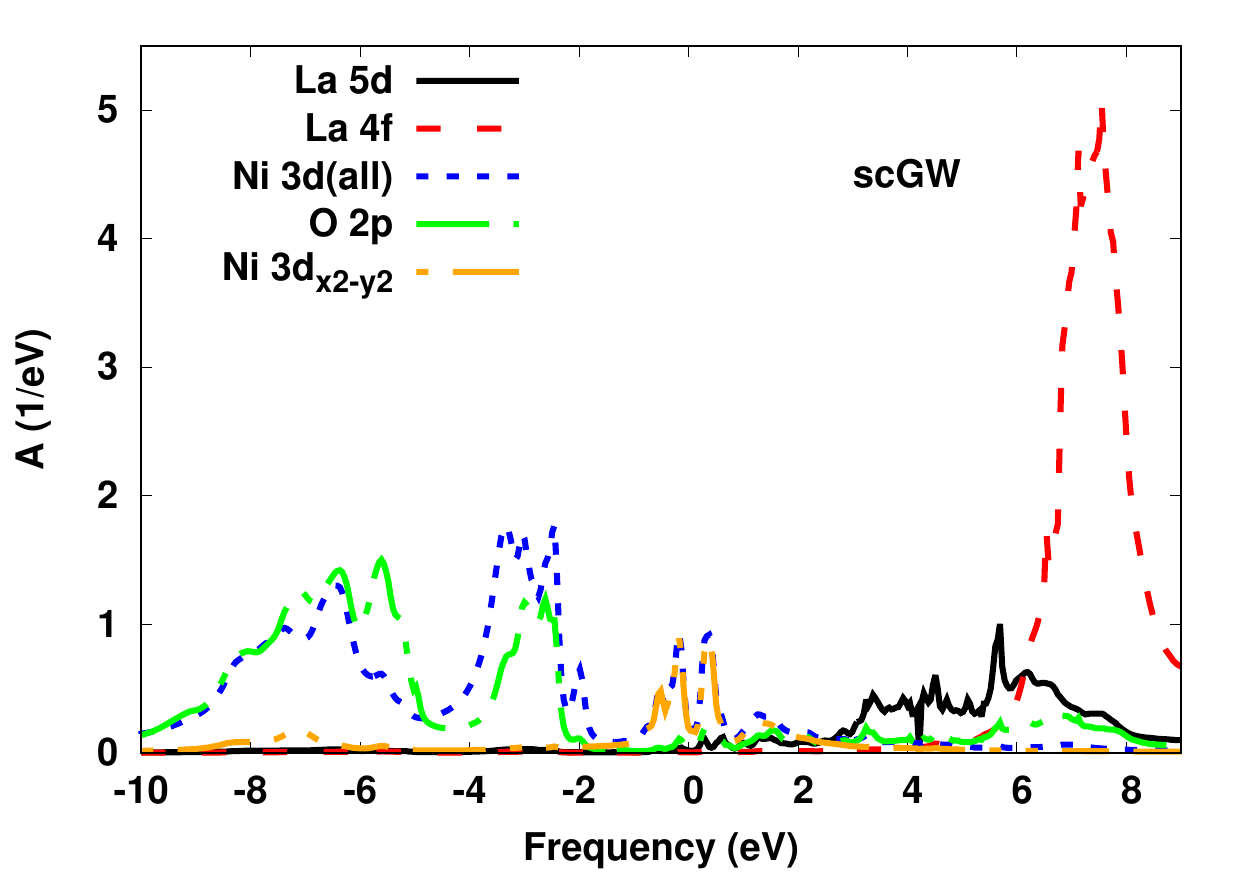}}  
    \hspace{0.02 cm}
    \fbox{\includegraphics[width=6.5 cm]{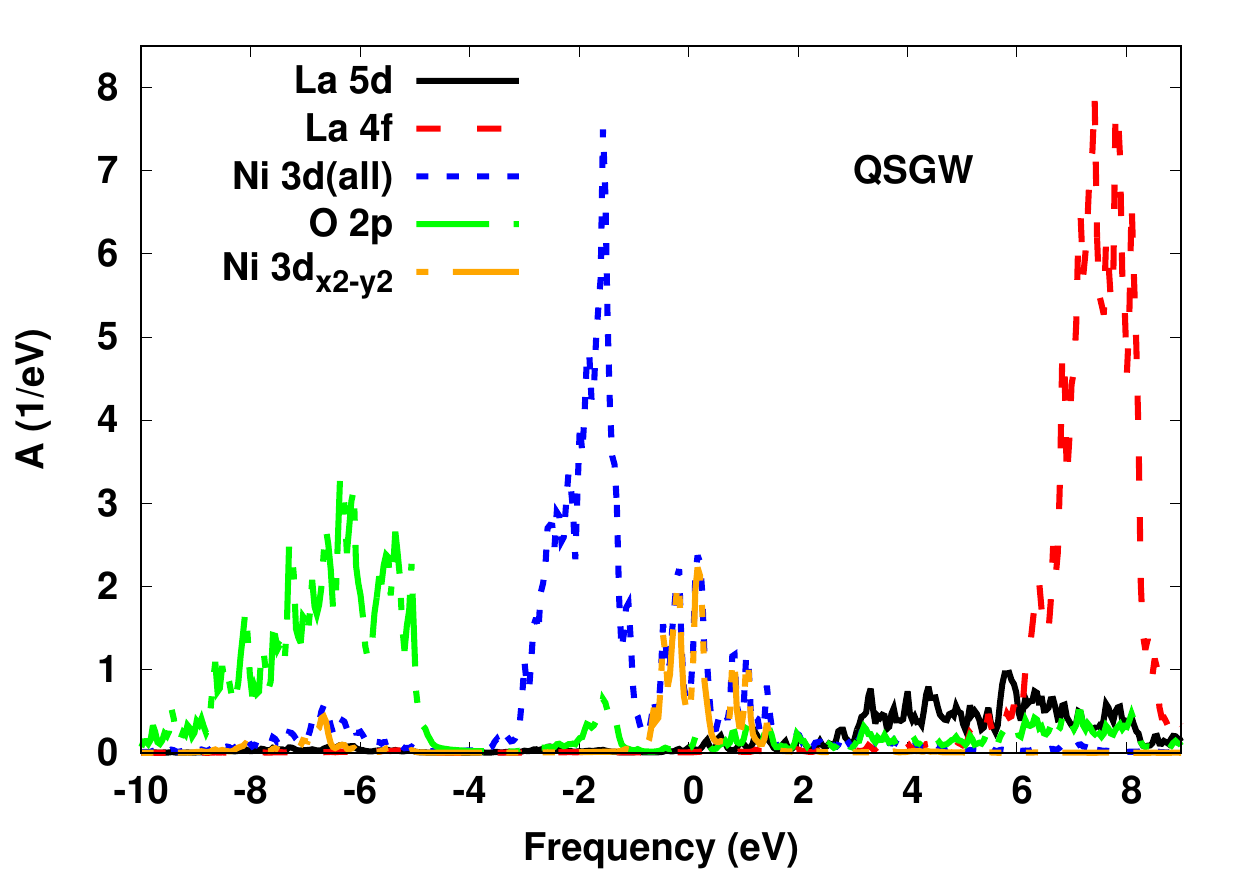}}  
    \hspace{0.02 cm}
    \fbox{\includegraphics[width=6.5 cm]{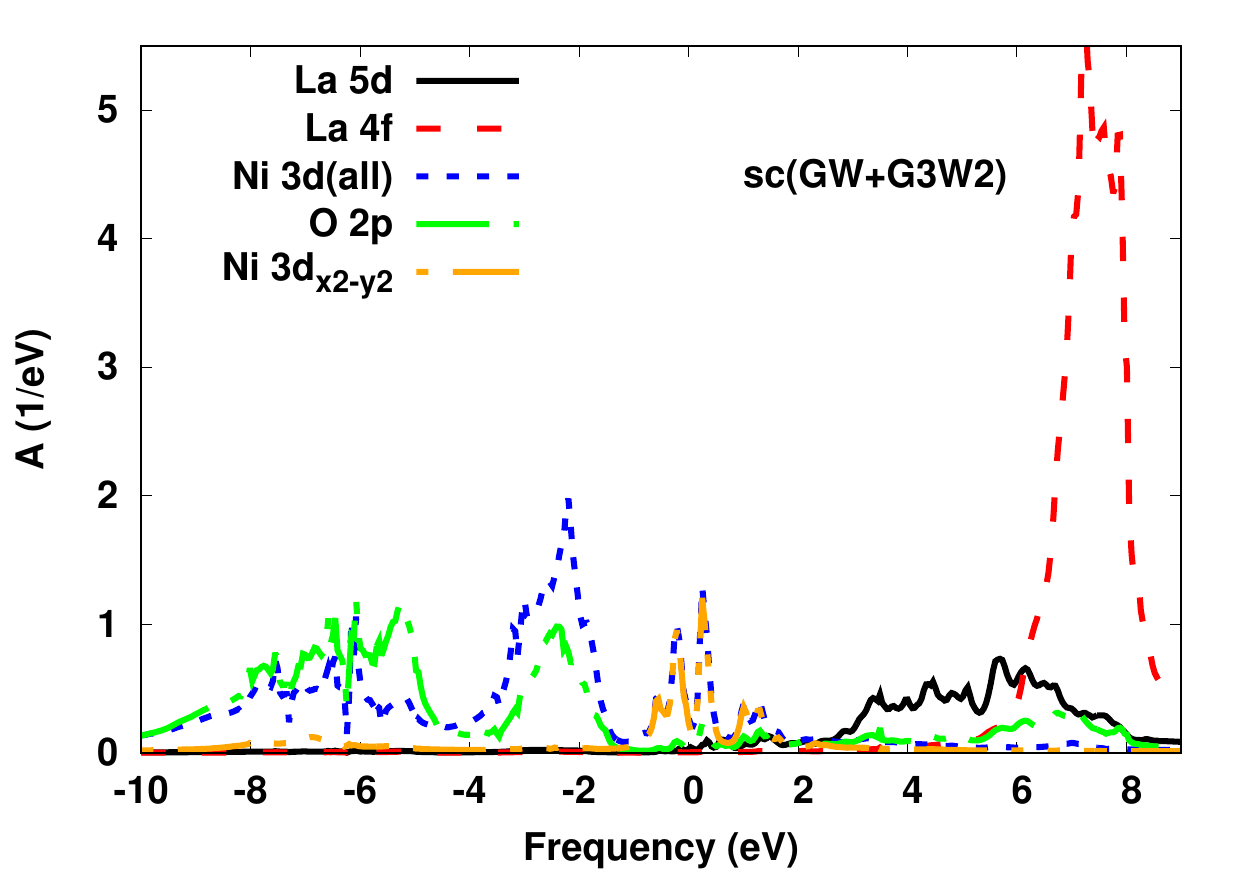}}
    \caption{Partial (atom and orbital resolved) spectral functions of LaNiO$_{2}$.}
    \label{pdos_la}
\end{figure*}

Partial (atom and orbital resolved) spectral functions are presented in Fig. \ref{pdos_ca} (CaCuO$_{2}$) and in Fig. \ref{pdos_la} (LaNiO$_{2}$). First, let us point out that there are a few important differences in the electronic structure of these two materials at the DFT level. First, the La 4f levels in LaNiO$_{2}$ dominate in the energy range immediately above the Fermi level. The La 5d states are spread in energy and are above of the 4f-states by 2-5 eV. Absence of the f-states in CaCuO$_{2}$ makes the presence of the Ca 3d states more prominent among the unoccupied bands. The character of the levels at the Fermi level also represents an important qualitative difference. In CaCuO$_{2}$, they are almost equally represented by the Cu 3d$_{x^{2}-y^{2}}$ and the O 2p states. In LaNiO$_{2}$, however, the 3d$_{x^{2}-y^{2}}$ states of Ni dominate. The states below the Fermi level also look different. In CaCuO$_{2}$, the Cu 3d states are mixed with the O 2p states and together they occupy the same energy range from -7eV to almost the Fermi level. In LaNiO$_{2}$, the Ni 3d states are well separated from the O 2p states and occupy the energy range from -3eV to the Fermi level, whereas the O 2p states occupy the energy range from -8eV to -3.5 eV.

Let us now discuss the changes in the electronic structure (as compared to the DFT) when we apply the fully self-consistent GW approach. In CaCuO$_{2}$, there are no qualitative changes. For instance, states at the Fermi level still are equally represented by the Cu 3d$_{x^{2}-y^{2}}$ and the O 2p states. Also, part of the spectral weight associated with 
the Cu 3d$_{x^{2}-y^{2}}$ states still resides in the occupied valence bands. The occupied Cu 3d states are strongly mixed with the O 2p states as in the DFT case. However, these joint occupied states are shifted down by about 2 eV as compared to the DFT case. The unoccupied Ca 3d states are shifted up by about 2 eV. In LaNiO$_{2}$, the states immediately at the Fermi level are still almost completely represented by the Ni 3d$_{x^{2}-y^{2}}$ orbitals, as in the DFT calculations. However, the rest of the electronic structure is qualitatively different from the DFT case. Firstly, the La 4f states are pushed up by about 5 eV in the scGW calculations as compared to the DFT. Now they are above of the La 5d bands and, supposedly, are not very important for the low energy physics. But even more noticeable change is related to the fact, that the occupied Ni 3d and the O 2p states which were very well separated in the DFT calculations now are strongly mixed and reside in the same energy range from approximatelu -10 ev to -2 ev relative to the Fermi level. As one can notice, this was achieved by a considerable down push of the occupied Ni 3d states and by a slight (about 1 eV) push of the O 2p states up in energy.

The self consistent vertex corrected GW calculations do not change the scGW result very much. One can notice, however, that in both materials the occupied Ni(Cu) 3d and O 2p states were pushed up in energy by about 0.5 eV (compared to the scGW result). Very slight down push of the Ca 3d (La 5d) can also be noticed.

The QSGW calculations for CaCuO$_{2}$ result in the electronic structure very similar to the electronic structure obtained with the scGW/sc(GW+G3W2) approach which can be verified by comparison of the positions of all principal peaks in Fig. \ref{pdos_ca}. The situation with LaNiO$_{2}$ is, however, quite different. Opposite to the scGW and the sc(GW+G3W2) calculations, the QSGW shows only quantitative (but not qualitative) changes in the electronic structure (as compared to the DFT). The only obvious similarity with the scGW results is the upward shift of the La 4f states. The oxygen 2p and the occupied Ni 3d states are pushed downward by -2 eV and -1 eV correspondingly, but there is no mixing among them as in the scGW or the sc(GW+G3W2) case. Obviously, in the case of LaNiO$_{2}$ the differences in methods, the QSGW on one hand and the scGW/sc(GW+G3W2) on the other hand, are a lot more prominent than in the case of CaCuO$_{2}$.

It is interesting to compare the tendencies in the electronic structure of LaNiO$_{2}$ (when we go from LDA to more complicated methods) with the tendencies discovered in Ref. [\onlinecite{prb_101_161102}]. Principal finding of Ref. [\onlinecite{prb_101_161102}] is that La 4f states are pushed up by about 2 eV in G0W0 calculation (as compared to the DFT), O 2p states are pushed down by about -1.5 eV, and the energy levels near the Fermi level do not change noticeably. Energy levels near the Fermi level are represented by Ni 3d$_{x^{2}-y^{2}}$) orbitals in all our calculations. In this respect, we agree with earlier G0W0 calculations. Further, all our post-DFT approaches push La 4f states up by about 5 eV which is larger than 2 eV in G0W0 case and can naturally be explained by self consistency effects. The change in the position of O 2p states is, however, different. Only our QSGW approach agrees with G0W0 finding: down push by about -1.5 eV. As it was already discussed above, the methods with dynamic self energy (scGW and sc(GW+G3W2)) demonstrate qualitatively different change: O 2p states split into two groups with the boundary between groups at about - 4 eV and mix considerably with Ni 3d states. The difference in this tendency is, most likely, related to the incoherence effects in self-consistency diagrams which are not included in G0W0 or QSGW approaches.

\begin{figure*}[t]       
    \fbox{\includegraphics[width=6.5 cm]{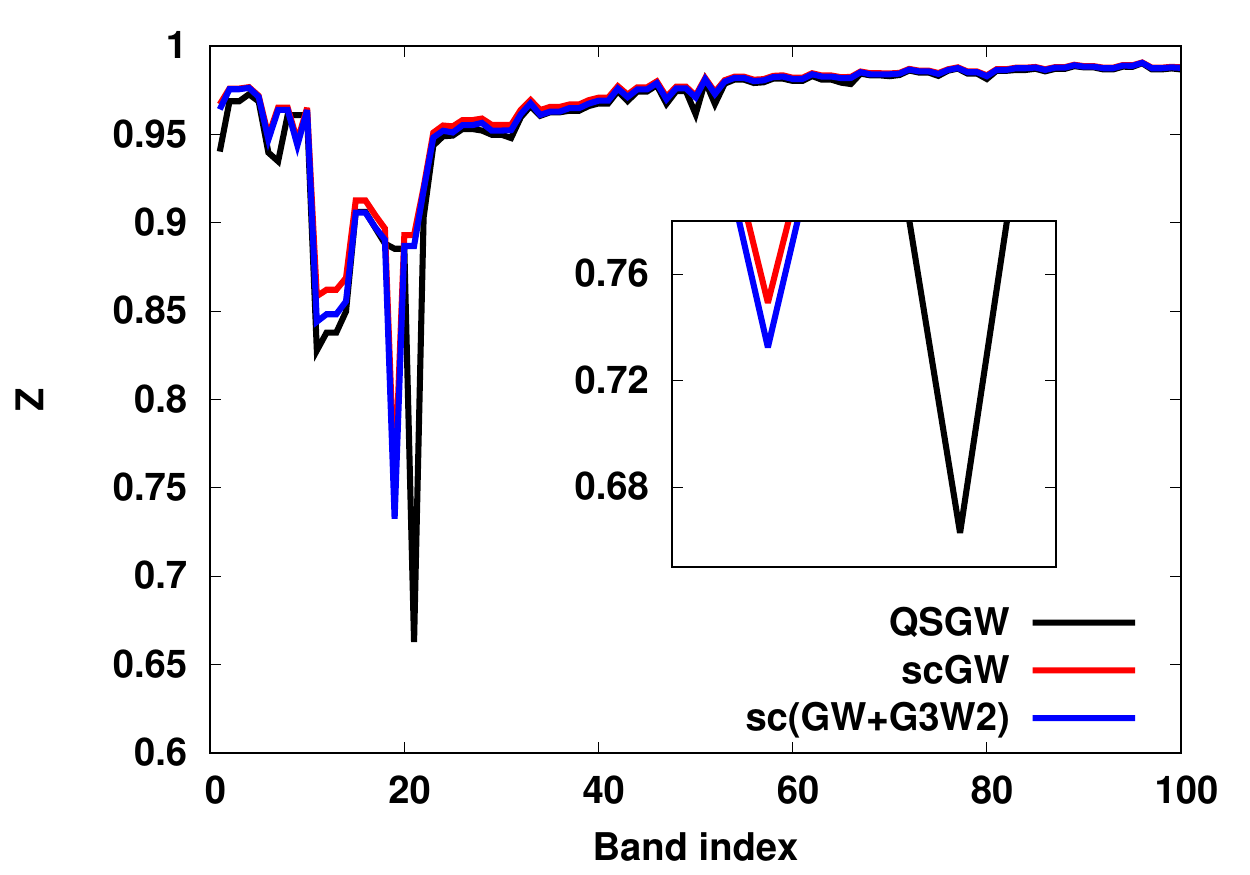}}   
    \hspace{0.02 cm}
    \fbox{\includegraphics[width=6.5 cm]{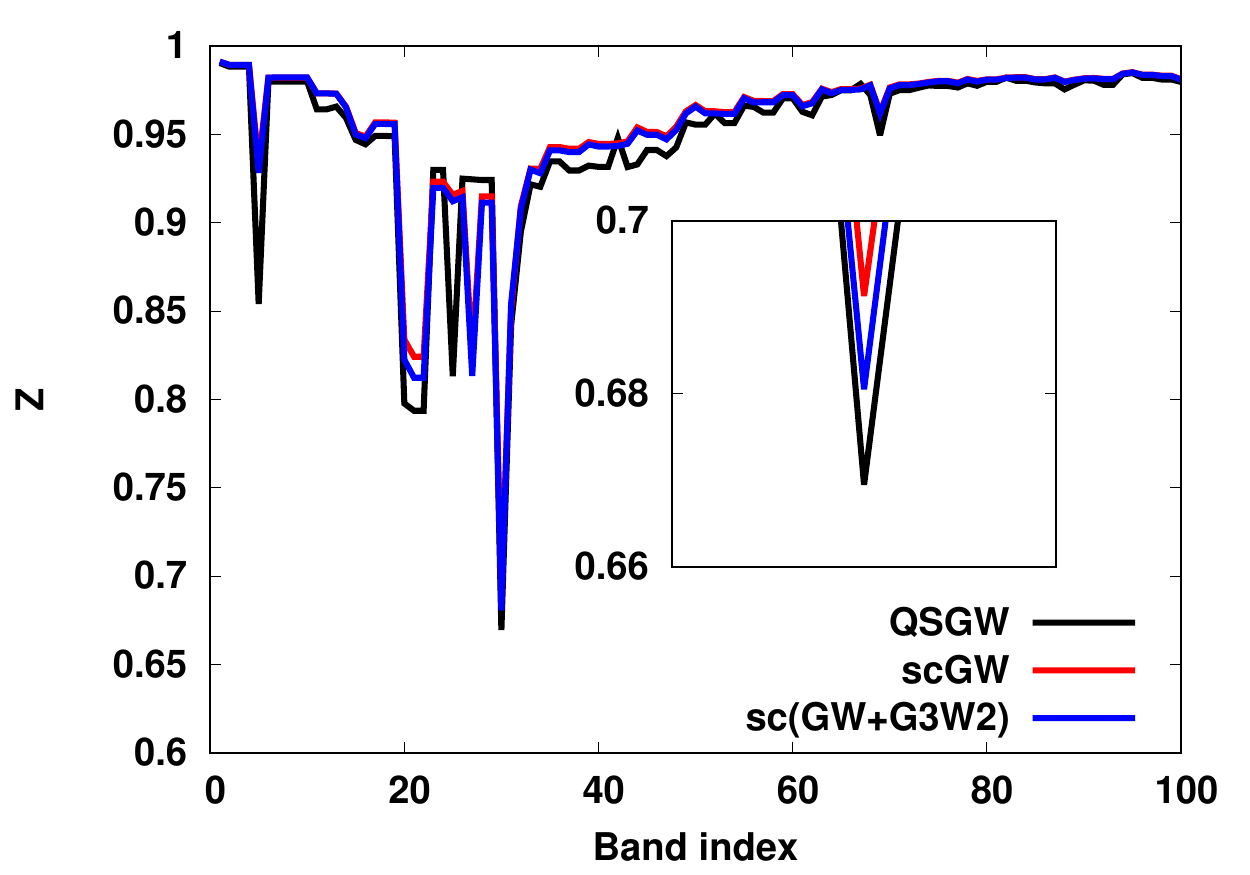}}
    \caption{Diagonal elements of the renormalization factor Z versus band index for $\Gamma$ point in the Brillouin zone. Left window: CaCuO$_{2}$, right window: LaNiO$_{2}$. The inserts magnify the area around the minimum of Z.}
    \label{z_factor}
\end{figure*}

In order to take measure of the strength of the correlation effects, the renormalization factor Z has been evaluated. Results are shown in Fig. \ref{z_factor} for the $\Gamma$ point of the Brillouin zone. As one can see, all approaches (scGW, sc(GW+G3W2), and QSGW) result in quite similar and moderate correlation effects. Minimal value of Z is unmistakebly obtained for one band near the Fermi level which has Ni(Cu) 3d$_{x^{2}-y^{2}}$ character. This holds for all points in the Brillouin zone in case of LaNiO$_{2}$. For CaCuO$_{2}$, however, there is noticeable admix of the O 2p character in some parts of the Brillouin zone (not shown in Fig. \ref{z_factor}). In those points, Z factor is slightly larger (up to 0.77$\div$0.80). Generally, analysis of Z factor confirms that the correlations are slightly stronger in LaNiO$_{2}$ case. In this respect, our calculations are in line with all published works. Also, similar to the G0W0 calculations\cite{prb_101_161102}, we obtained very little variation of Z across the Brillouin zone for the Ni 3d$_{x^{2}-y^{2}}$ band in case of LaNiO$_{2}$. Its value also is very close to the value $0.70\pm0.02$ reported in the G0W0 calculations. There are, however, differences with the DFT+DMFT results. Most notable is that Z factor in the DFT+DMFT calculations\cite{prx_10_021061} is considerably smaller for the most correlated Ni 3d$_{x^{2}-y^{2}}$ orbital. Its value was reported to be 0.36 (LaNiO$_{2}$, Ref. \onlinecite{prb_102_161118}). In the case of the Cu 3d$_{x^{2}-y^{2}}$ orbital, the DFT+DMFT values are 0.50$\pm$0.75 (CaCuO$_{2}$, Ref. \onlinecite{prx_10_021061}) which are not much different from ours. There are a few possible sources of the differences for LaNiO$_{2}$: i) insufficient number of diagrams included in our calculations; ii) single site approximation in the DFT+DMFT calculations; iii) the Hubbard U was taken too large in the DFT+DMFT case. Experimental research is, therefore, imperative for the purpose of comparison. However, as all calculations neglect the electron-phonon interaction, direct comparison with future experimental mass enhancement (for instance) will require inclusion of the electron-phonon interaction in the theoretical predictions.

\begin{figure*}[t]       
    \fbox{\includegraphics[width=6.5 cm]{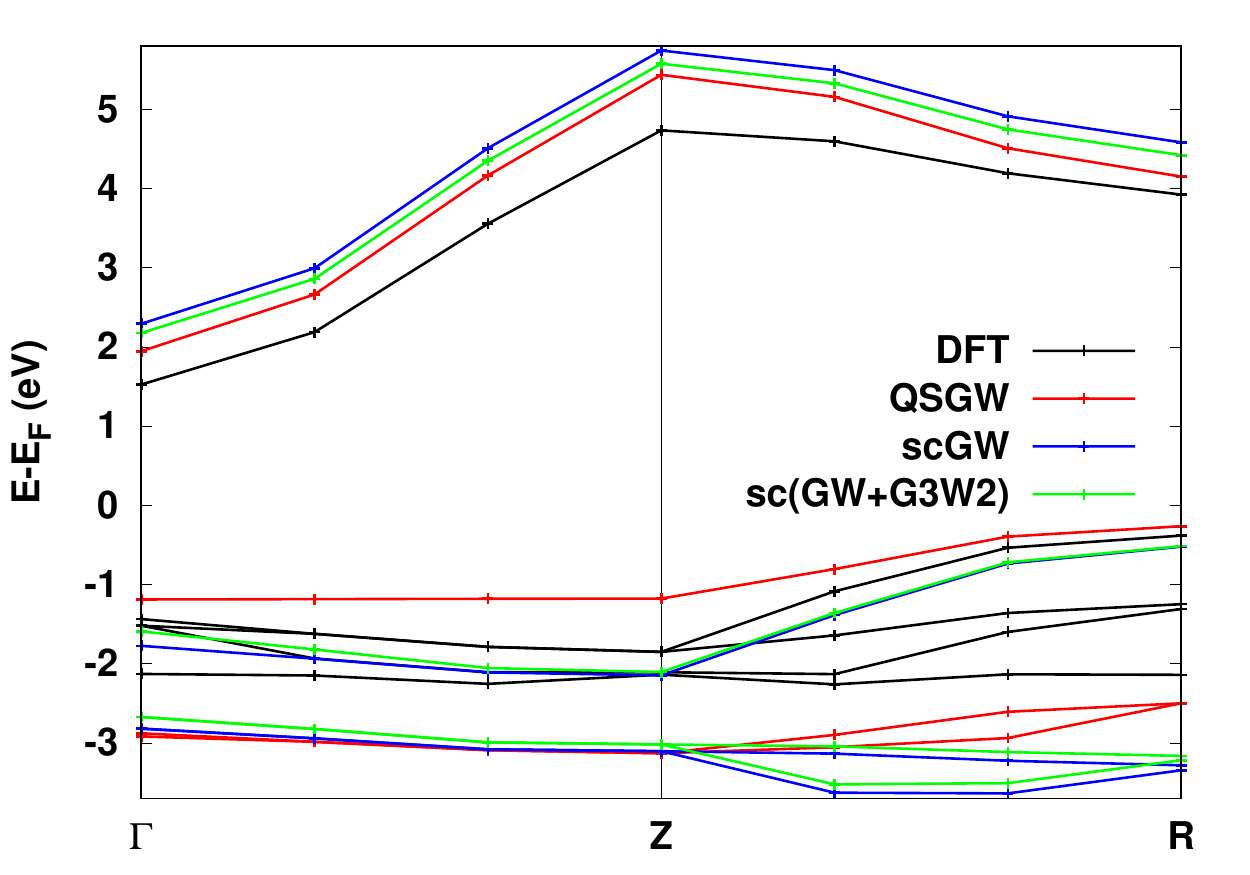}}   
    \hspace{0.02 cm}
    \fbox{\includegraphics[width=6.5 cm]{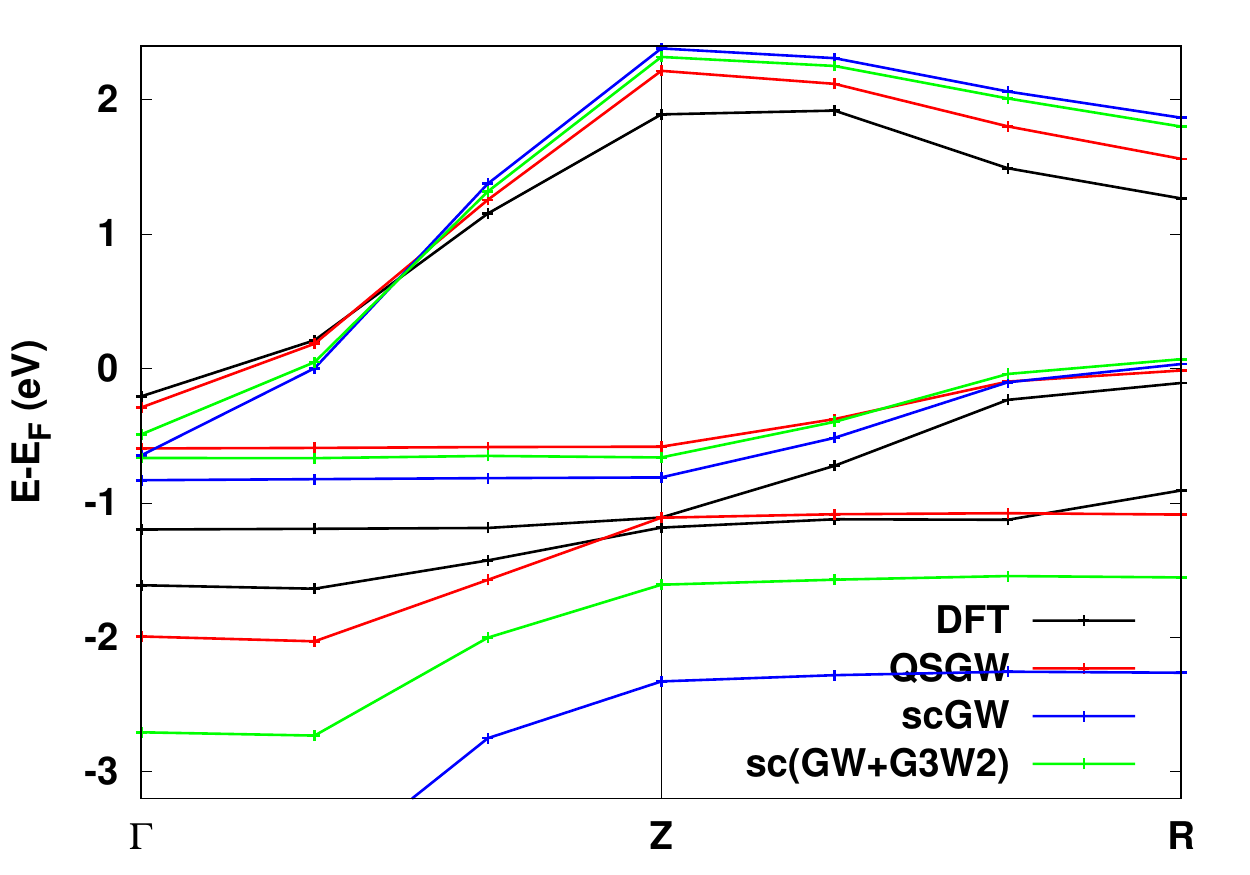}}
    \caption{Quasiparticle band structure along $\Gamma$-Z-R path in the Brillouin zone. Left window: CaCuO$_{2}$, right window: LaNiO$_{2}$. Only the bands near Fermi level are shown.}
    \label{gzr_bnd}
\end{figure*}

Distinctive feature of the DFT+DMFT calculations is the renormalization (narrowing) of the bands near the Fermi level. For LaNiO$_{2}$ it is shown in Fig. 1 in Ref. \onlinecite{prb_102_161118} and for CaCuO$_{2}$ it is shown in Fig. 4 in Ref. \onlinecite{prx_10_021061}. Particularly well the narrowing is seen along the $\Gamma$-Z direction in the Brillouin zone for the band immediately under the Fermi level (LaNiO$_{2}$). In CaCuO$_{2}$ case, the actual bands are well entangled near the Fermi level, so that in the DMFT applications the disentaglement procedure is used. For the purpose of comparison, we show in Fig. \ref{gzr_bnd} the bands near the Fermi level along the path $\Gamma$-Z-R in the Brillouin zone. The case of LaNiO$_{2}$ is a bit simpler, so we discuss it first. As one can see at the $\Gamma$ point, all three correlated methods (QSGW, scGW, and sc(GW+G3W2)) result in large narrowing of the DFT band of the Ni 3d$_{x^{2}-y^{2}}$ character which is the second band from the Fermi level at $\Gamma$ point. This band can be easily identified: it starts at -1.2 eV in the DFT case, and at about -0.5 eV in the other cases. The strongest renormalization is in the sc(GW+G3W2) case (more than by a factor of two) which is approximately the same renormalization as in the DFT+DMFT case.\cite{prb_102_161118} The QSGW and the scGW result in only slightly smaller renormalization. The similarity with the DFT+DMFT result is interesting considering the number of differences in the methods. In the G0W0 calculations\cite{prb_101_161102} the band narrowing along the $\Gamma$-Z path was also observed but almost by a factor of two smaller than in our calculations because of lack of self consistency. If we look at the same band along the $\Gamma$-Z path, we can see the difference between methods. The band is flat in the DFT, the DFT+DMFT, and in the QSGW cases but it has a slight but noticeable dispersion in the scGW and in the sc(GW+G3W2). In case of the DFT+DMFT, it is flat because the DMFT self energy is independent of momenta and, correspondingly, the flatness of the DFT band remains. But the difference between the QSGW on one hand and the scGW/sc(GW+G3W2) on the other hand deserves attention. Taking into account the specifics of the methods, one can speculate that dynamic effects (frequency dependence) in self energy which is included in the scGW and in the sc(GW+G3W2) but not in the QSGW is the reason. Self energy is momentum-dependent in all three methods, so the difference in its frequency dependence from one \textbf{k}-point to another can result in dispersion. In the DFT, this band is crossing with another band at Z point. But there is no such crossing in any of the correlated methods including the DFT+DMFT. Thus, this is another similarity of our results with the DFT+DMFT results.

There is interesting difference with the DFT+DMFT in the size of the electron pocket near the $\Gamma$ point. In the DFT+DMFT,\cite{prb_102_161118} its size is slightly reduced as compared to the DFT case. In our calculations, all three correlated methods show an increase of the pocket. Slight increase of the electron pocket at $\Gamma$ was also reported in the G0W0 calculations.\cite{prb_101_161102} This would be interesting to compare with experiment. However, as it represents one of the low energy effects, the electron-phonon interaction has to be taken into account in the theoretical evaluations for proper comparison.

In CaCuO$_{2}$ case, the comparison is more complicated because of entanglement of the bands (O 2p orbitals contribute significantly). The band of interest (Cu 3d$_{x^{2}-y^{2}}$) is the fourth band (down from the Fermi level) and starts at the $\Gamma$ point at about -2.1 eV in the DFT case. In the correlated methods, however, this band is the first one down from the Fermi level and starts at about -1.5$\div$-1.8 eV (scGW and sc(GW+G3W2)) and at about -1.2 eV in the QSGW case. So, the renormalization is smaller (scGW and sc(GW+G3W2)) than in the DMFT case (Fig. 4 in Ref. \onlinecite{prx_10_021061}). Interesting, however, that in this case the narrowing is the strongest in the QSGW case which is close enough to the DFT+DMFT result. Once again, we need experimental information in order to decide which method is the best. If the QSGW is more accurate than we should conclude that long range static correlation effects are more important for this material than dynamic effects. If, however, the scGW/sc(GW+G3W2) is more accurate the conclusion would be opposite.


\section*{Conclusions}
\label{concl}

In conclusion, we have applied three correlated methods (scGW, sc(GW+G3W2), and QSGW) to study the electronic structure of CaCuO$_{2}$ and LaNiO$_{2}$. In some aspects, our results are consistent with the previous DFT+DMFT studies: band narrowing near the Fermi level, orbital differentiation in the Ni(Cu) 3d shell, stronger correlation effects in LaNiO$_{2}$ as compared to CaCuO$_{2}$. There are also differences with the DFT+DMFT studies. One of them consists in quite noticeable repositioning of the spectral features away from the Fermi level in our correlated calculations. In the DMFT case, repositioning is small because only the Ni(Cu) 3d electrons are considered as correlated. Another notable difference consists in a lot smaller Z factor obtained in the DFT+DMFT works for LaNiO$_{2}$. This could be a result of the insufficient number of diagrams in our calculations, or simply the artifact of the single-site approximation and/or the too large Hubbard U parameter in the DFT+DMFT studies. Also, the change in the size of the electron pocket near the $\Gamma$ point is different: small decrease in the DMFT case and an increase in all our GW based methods. In this respect, out GW based approaches agree with the result found in Ref. [\onlinecite{prb_101_161102}] using G0W0 approximation. Concerning the above mentioned repositioning of the Ni 3d and O 2p levels below the Fermi level, we have found similarity with G0W0 calculations\cite{prb_101_161102} in our QSGW studies, but not in our scGW or sc(GW+G3W2) studies. We have also found that our three correlated methods differ between each other more prominently in the case of LaNIO$_{2}$ which is consistent with the conclusion that this material is more correlated.

Principal results of this work show that the correlations in LaNiO$_{2}$ being stronger than in CaCuO$_{2}$ still, however, are weak enough to allow applications of the totally ab-initio methods such as the scGW or the more advanced sc(GW+G3W2) for both materials. Whether there is a physics which cannot be captured by perturbative methods like scGW or sc(GW+G3W2) and, therefore, one needs to use non-perturbative approaches like DFT+DMFT, still has to be explored by future photoemission experiments.

\section*{Acknowledgments}
\label{acknow}
This work was   supported by the U.S. Department of energy, Office of Science, Basic
Energy Sciences as a part of the Computational Materials Science Program.

\appendix

\section{Details of the evaluation of some diagrams}\label{diag_det}

\begin{figure}[t]
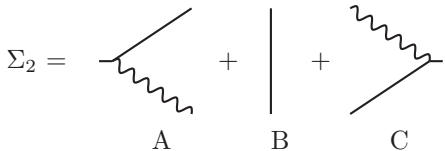

\begin{center}\begin{axopicture}(200,56)(0,-10)
\SetPFont{Arial-bold}{28}
\SetWidth{0.8}
\Text(10,10)[l]{$\Sigma_{2}$  =}
\Text(65,-20)[l]{A}
\Text(110,-20)[l]{B}
\Text(155,-20)[l]{C}
\Line(45,10)(50,10)
\Photon(50,10)(80,-10){2}{5.5}
\Line(50,10)(80,30)
\Text(90,10)[l]{$+$}
\Line(110,-10)(110,30)
\Text(125,10)[l]{$+$}
\Photon(140,30)(170,10){2}{5.5}
\Line(140,-10)(170,10)
\Line(170,10)(175,10)
\end{axopicture}
\end{center}
\caption{Scheme of the evaluation of the second order diagram for self energy.}
\label{sigma_steps}
\end{figure}

\begin{figure}[t]
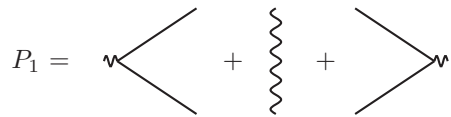

\begin{center}\begin{axopicture}(200,56)(0,-10)
\SetPFont{Arial-bold}{28}
\SetWidth{0.8}
\Text(10,10)[l]{$P_{1}$  =}
\Photon(45,10)(50,10){2}{1.5}
\Line(50,10)(80,-10)
\Line(50,10)(80,30)
\Text(90,10)[l]{$+$}
\Photon(110,-10)(110,30){2}{5.5}
\Text(125,10)[l]{$+$}
\Line(140,30)(170,10)
\Line(140,-10)(170,10)
\Photon(170,10)(175,10){2}{1.5}
\end{axopicture}
\end{center}
\caption{Scheme of the evaluation of the first order diagram for irreducible polarizability.}
\label{p_steps}
\end{figure}

In Ref. [\onlinecite{prb_94_155101}] the evaluation of the vertex correction to self energy was presented as a two-step process. In the first step, the non-trivial part of three-point vertex function $\Gamma$ was evaluated. In the second step, this vertex function was combined with Green's function G and screened interaction W to form self energy $GW\Gamma$. However, it was found later, that a considerably more efficient procedure for the evaluation of second order self energy consists in the evaluation of the corresponding diagram directly, avoiding the intermediate construction of $\Gamma$. Namely, the second diagram presented in Fig. \ref{diag_S} can be evaluated in three steps as it is demonstrated in Fig. \ref{sigma_steps}. The pieces A, B, and C shown in Fig. \ref{sigma_steps} are combined together beginning from the right and proceeding to the left. Essentially, the algorithm is very similar to the algorithm for first order polarizability shown schematically in Fig. \ref{p_steps} and described in details in Ref. [\onlinecite{prb_94_155101}].

The piece C (Fig. \ref{sigma_steps}) is evaluated in (reciprocal space + frequency) representation with the band states indexes representing the orbital basis set. After evaluation, piece C is transformed into (real space + imaginary time) representation (see Ref. [\onlinecite{prb_94_155101}] for the specifics of the representations of functions in real space). Thus, pieces B and C are combined in (real space + imaginary time) representation, which approximately can be thought of as point by point multiplication. After that, the object B+C is transformed back to the (reciprocal space + frequency) representation, and it is combined with the piece A. In practice, this algorithm of self energy evaluation is a few times faster than the original one presented in Ref. [\onlinecite{prb_94_155101}] and requires considerably less memory.

\section{Effective correlation self energy in QSGW}\label{sig_qsgw}

For clarity we define effective correlation self energy $\Sigma^{corr}$ in QSGW as self energy which transforms Hartree-Fock Green's function $G^{HF}$ into QSGW Green's function $G^{QP}$ via Dyson's equation:

\begin{equation}\label{dyson}
\Sigma^{corr}=G^{^{-1}HF}-G^{^{-1}QP}.
\end{equation}

In the above equation, we express all quantities in the basis of the Hartree-Fock eigen-states $\Psi^{\mathbf{k},HF}_{\lambda}(\mathbf{r})$, where \textbf{k} is momentum and $\lambda$ stands for the Hartree-Fock band state. In this basis set, Hartree-Fock Green's function is diagonal:

\begin{equation}\label{g_hf}
G^{\mathbf{k},HF}_{\lambda\lambda'}(\omega)=\frac{\delta_{\lambda\lambda'}}{i\omega+\mu-\epsilon^{\mathbf{k}}_{\lambda}},
\end{equation}
where $\omega$ is Matsubara's frequency, $\epsilon^{\mathbf{k}}_{\lambda}$ is the Hartree-Fock one electron energies, and $\mu$ stands for the chemical potential. QSGW Green's function in the basis of Hartree-Fock band states has the following form (see Section 5 in Ref. [\onlinecite{cpc_219_407}]):

\begin{align}  \label{g_qp}
G^{\mathbf{k}, QP}_{\lambda\lambda'}(\omega)= 
\sum_{i}\frac{Q^{\mathbf{k}}_{\lambda i}Q^{^{\dagger}\mathbf{k}}_{i\lambda'}}{i\omega+\mu-E^{\mathbf{k}}_{i}},
\end{align}
with $E^{\mathbf{k}}_{i}$ being quasiparticle one electron energies and with unitary matrices $Q^{\mathbf{k}}_{\lambda i}$ representing transformation from Hartree-Fock states to QSGW (QP) states $\Psi^{\mathbf{k},QP}_{i}(\mathbf{r})=\sum_{\lambda}Q^{\mathbf{k}}_{\lambda i}\Psi^{\mathbf{k},HF}_{\lambda}(\mathbf{r})$.

Direct substitution of (\ref{g_hf}) and (\ref{g_qp}) in eq. (\ref{dyson}) gives simple result:

\begin{align}  \label{sig_qp}
\Sigma^{\mathbf{k}, corr}_{\lambda\lambda'}= 
\sum_{i}Q^{\mathbf{k}}_{\lambda i}E^{\mathbf{k}}_{i}Q^{^{\dagger}\mathbf{k}}_{i\lambda'}-\epsilon^{\mathbf{k}}_{\lambda}\delta_{\lambda\lambda'}.
\end{align}

\begin{figure*}[t]       
    \fbox{\includegraphics[width=6.5 cm]{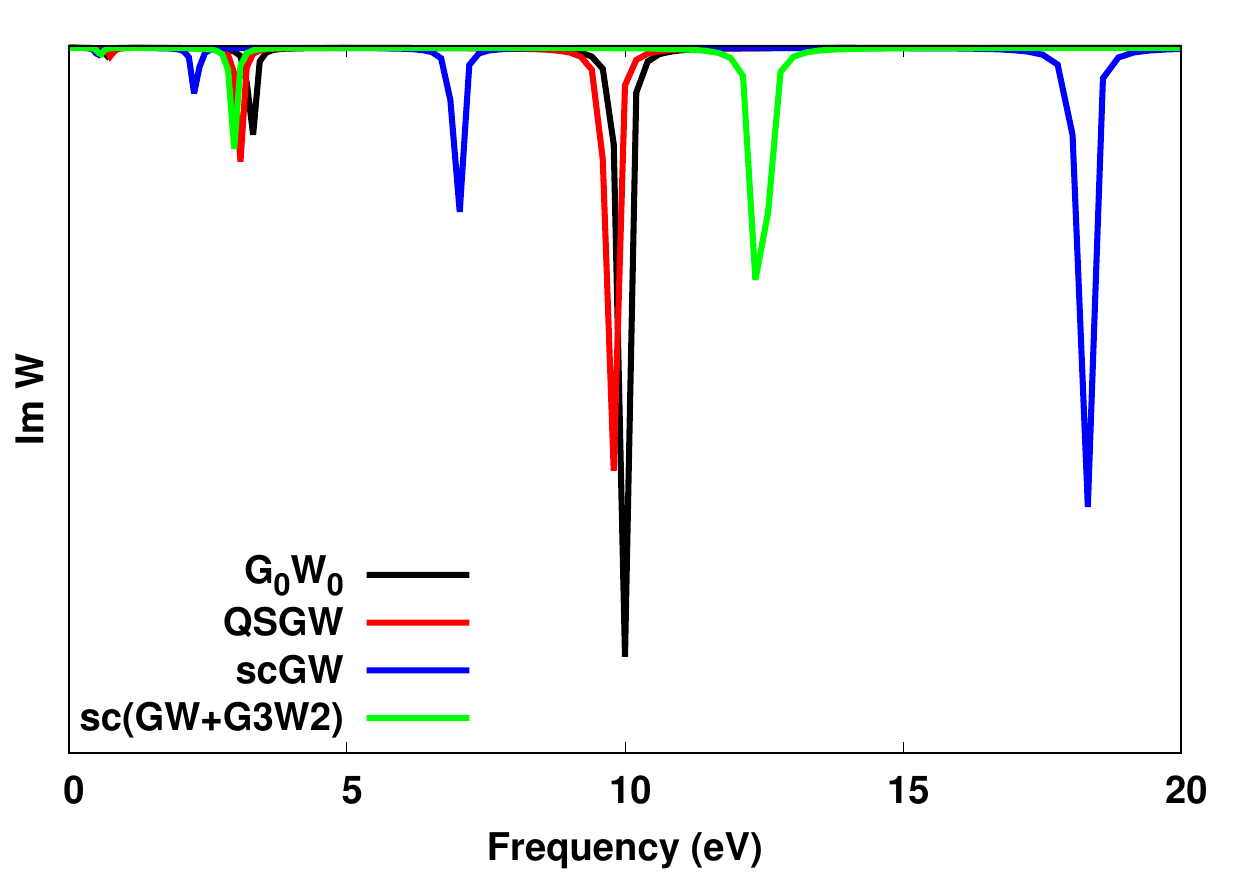}}   
    \hspace{0.02 cm}
    \fbox{\includegraphics[width=6.5 cm]{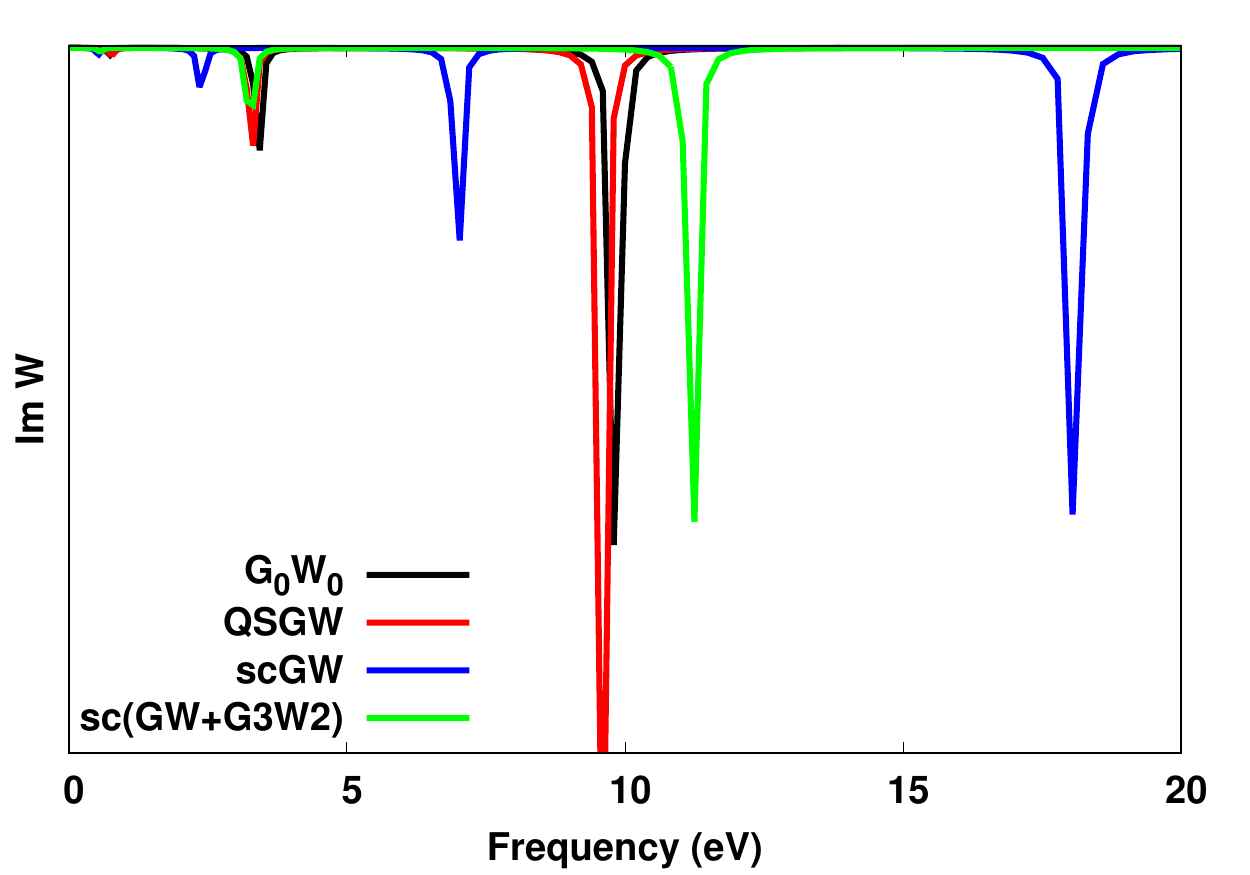}}
    \caption{Diagonal elements of the Im W evaluated for Ni 3d(yz)(left window) and for Ni 3d(3z$^{2}$-r$^{2}$)(right window) orbitals as functions of frequency (real axis). The results are for LaNiO$_{2}$.}
    \label{imw}
\end{figure*}

As one can see, effective correlation self energy in QSGW is explicitly frequency independent (static). However, as it also is clear from the construction\cite{cpc_219_407} of the QSGW Green's function, there is implicit dependence of its matrix elements on the frequency. This implicit frequency dependence comes from the fact that quasiparticle energies $E^{\mathbf{k}}_{i}$ are obtained after the linearization of the frequency dependence of the "original" diagrammatic self energy $\Sigma=GW$. Any change in frequency dependence of the "original" self energy would result in a change of the quasiparticle energies $E^{\mathbf{k}}_{i}$ and, as a result, in a change of the effective correlation self energy.

The absence of explicit dependence of the QSGW self energy on frequency creates certain qualitative differences between QSGW and self-consistent diagrammatic approaches (like scGW). For instance, there are no incoherent effects in QSGW, and its one electron energies are well defined (i.e. have infinite lifetime) like one electron energies in DFT or in Hartree-Fock approximations.

\section{Fully screened interaction as a function of real frequency}\label{Im_W}

In order to shed a little bit more light on the differences between methods based on the well defined quasiparticles (QSGW) and methods with incoherent effects (scGW and sc(GW+G3W2)) we plot imaginary part of the screened interaction W (Im W) as a function of real frequency (Fig. \ref{imw}). The results obtained in RPA with LDA Green's function are also included (abbreviated as G0W0). In FlapwMBPT code, everything is done in Matsubara's formalism. Therefore, in order to plot W as a function of real frequency, it has to be analytically continued from imaginary to real axis of frequencies. This was done with use of the algorithm proposed for  analytical continuation (AC) of bosonic functions by Vidberg and Serene in Ref.  [\onlinecite{jltp_29_179}]. We have to mention that for the analytical continuation of self energy (fermionic function) we use similar algorithm\cite{prb_85_155129} which is a slight modification of the algorithm by Vidberg and Serene. There is, however, a certain difference in the degree of robustness of the AC for fermionic and bosonic functions. Whereas AC of self energy is quite robust which was tested on numerous materials (we do it all the time when calculate  electronic structure), the result of AC of bosonic functions is more sensitive to the quality of input information on the imaginary axis. It was noticed a few years before when calculating the electron energy loss spectrum (EELS) of LiF.\cite{prb_95_195120} In this work, we also had to double the number of points on both imaginary time and imaginary frequency grids (from 64 to 128) in order to stabilize the positions of peaks of Im W within 0.5-1 eV. Nevertheless, qualitative picture (relative positions of peaks obtained by different methods) didn't change much during the process. With the above information kept in mind, we can notice that there is clear difference between G0W0, QSGW and sc(GW+G3W2) on one hand and scGW on the other hand. As compared to the approaches from the first group which have only one peak in the interval 5-20 eV, scGW results in two peaks in the indicated interval of frequencies. Taking into account the fact that scGW is, most likely, the less accurate of the studied approaches, one can assume that the peak at about 7 eV obtained in scGW is a result of approximations. It is hardly possible to ascribe precise physical meaning to the obtained peaks, partly because of subtleties of AC and also because of the complexity of the electronic structure of LaNiO$_{2}$. The peaks below 5 eV are most likely the artifacts of the AC as their amplitude reduces when we increase accuracy of the input data (W on the imaginary axis). The peaks above 5 eV are most likely of plasmon nature and, in this respect, the situation seems to be different from, for instance, the electron gas. In electron gas,\cite{prb_57_2108} the plasmon poles obtained with G0W0 are flushed out when one uses scGW. In LaNiO$_{2}$, on the other hand, there are two poles in scGW instead of one pole in G0W0 or QSGW. When one uses sc(GW+G3W2) approximation, the number of poles in the interval 5-20 eV becomes one, and the pole is shifted towards smaller frequencies as compared to the case of scGW. In this respect, the situation reminds the one in LiF\cite{prb_95_195120} where the positions of poles of the dielectric function are also shifted towards smaller frequencies when one uses vertex-corrected scheme instead of scGW. In general, three methods (G0W0, QSGW, and sc(GW+G3W2)) seem to be consistent between each other whereas scGW stands a bit apart from them. 


\end{document}